\documentclass[11pt]{article}

\usepackage{moriond}
\usepackage{epsfig}

\title{Bayesian Study and Naturalness in MSSM Forecast for the LHC}

\author{Maria Eugenia Cabrera}

\address{Instituto de F\'isica Te\'orica, IFT-UAM/CSIC \\ 
         U.A.M., Cantoblanco, \\ 
         28049 Madrid, Spain}

\begin{document}

\begin{flushright}
{\small
IFT-UAM/CSIC-10-36\\}
\end{flushright}

\vspace*{1.5cm}

\maketitle

\abstracts{We perform a forecast of the CMSSM for the LHC based
  in an improved Bayesian analysis taking into account the present
  theoretical and experimental wisdom about the model
  \cite{Cabrera:2009dm}. In this way we obtain a map of the preferred
  regions of the CMSSM parameter space and show that fine-tuning
  penalization arises from the Bayesian analysis itself when the
  experimental value of $M_Z$ is considered. The results are
  remarkable stable when using different priors.}

The start of the LHC has motivated a lot of effort to try to
anticipate which kind of physics beyond the Standard Model is more
likely to be there.  Since the present experimental data are not
powerful enough to select a small region of the parameter space of
SUSY models, Bayesian Statistics becomes a very powerful tool to try
to make an inference of the probability of certain regions of
parameters of these models, where the choice of judicious prior
probability for the parameters becomes more relevant.

\section{Bayesian Statistics}

The probability density of a particular
point $\{p_i^0\}$ in the parameter space given a certain set of
\textit{data} is the posterior probability density function
($\mathrm{pdf}$), given by
\begin{eqnarray}
\label{bayesTheorem}
p(p_i^0| \mathrm{data} ) = \frac{p(\mathrm{data}|p_i^0)\,p(p_i^0)}{p(\mathrm{data})},
\end{eqnarray}
where $p(\mathrm{data}|p_i^0)$ is the likelihood, the probability
density of measuring the given data for the chosen point in the
parameter space.  $p(p_i^0)$ is the prior, the ``theoretical''
probability density that we assign a priory to the point in the
parameter space. $p(\mathrm{data})$ is the evidence. If one is
interested in comparing regions of the parameter space of a given
model, $p(\mathrm{data})$ is just a normalization constant.

\section{Bayesian approach and Naturalness}

The parameters of the MSSM should not be far from the electroweak
scale in order to avoid unnatural fine-tunings to obtain the correct
scale of the electroweak breaking. From the minimization of the
tree-level form of the scalar potential,
\begin{eqnarray}
\label{MZ}
M_Z^2 &=& 2\ \frac{m_{H_1}^2 - m_{H_2}^2\tan^2 \beta}{\tan^2\beta-1}
-2\mu_{low}^2\ ,
\end{eqnarray}
we can see that if $\mu$ and the soft masses $m_{H_{1,2}}$ are not close
to the electroweak scale, a big cancellation is necessary in order to
obtain the right value of $M_Z$. A conventional measure of this
cancellation are the fine-tuning parameters \cite{Ellis:1986yg,Barbieri:1987fn},
\begin{eqnarray}
\label{BG}
c_i = \left|\frac{\partial \ln M_Z^2}{\partial \ln p_i}\right|.
\end{eqnarray}
Since naturalness arguments are actually statistical, one may expect
that a penalization of fine-tunings should arise from the Bayesian
analysis itself. Let us see how this comes about. Let us consider
$M_Z$ on a similar foot to the rest of experimental data,
\begin{eqnarray}
\label{likelihood}
p({\rm data}|s, m, M, A, B, \mu)\ = {\cal L}_{M_Z}\ {\cal L}_{\rm rest}\ ,
\end{eqnarray}
where $s$ represents the SM parameters, $\cal{L}_{\rm{rest}}$ is the
likelihood of the all physical observables except $M_Z$, and ${\cal
  L}_{M_Z}$ is the likelihood of $M_Z$. Let us now use the sharpness
of the likelihood of $M_Z$ to approximate ${\cal{L}}_{M_Z} \simeq
\delta(M_z - M_z^{exp})$ and marginalise the pdf in the
$\mu-$parameter, performing a change of variable $\mu\rightarrow M_Z$:
\begin{eqnarray}
\label{marg_mu}
p(s, m, M, A, B| \ {\rm data} ) &\simeq& \int dM_Z \left[\frac{d\mu}{d
    M_Z}\right] \delta(M_z - M_z^{exp})\, {\cal L}_{\rm rest}\ p(s, m,
M, A, B, \mu) \nonumber \\ 
&=& \left[\frac{d\mu}{d
    M_Z}\right]_{\mu_0} {\cal L}_{\rm rest} \; p(s, m, M, A, B,
\mu_0)\ ,
\end{eqnarray}
where $\mu_0$ is the one which reproduce the experimental value of
$M_Z$ as a function of the rest of the parameters \footnote{We have
  ignored here the nomalization factor. See
  eq. [\ref{bayesTheorem}]}. Now, comparing this to the definition of
fine-tuning parameters one gets,
\begin{eqnarray}
\label{pcmu}
 p(s, m, M, A, B| \ {\rm data} )\
 = \ 2 \frac{\mu_0}{M_Z}\
 \frac{1}{c_\mu}\ {\cal L}_{\rm rest}\; p(s, m, M, A, B, \mu_0),
\end{eqnarray}
where the presence of the fine-tuning parameter does indeed penalize
regions of the parameter space with large fine-tunings. As we will
see, this is enough to make the high-energy region of the parameter
space statistically insignificant.

We have
performed a Bayesian analysis of the MSSM with the following
improvements \cite{Cabrera:2008tj}: as we show above, the fine-tuning
penalization arises from the Bayesian analysis itself; we have made a
rigorous treatment of the nuisance variables where Yukawa couplings
are fundamental parameters in contrast with previous analysis; and,
finally, we have used an efficient set of variables to scan the MSSM,
$\{m,M,A,B,mu,y_t\}\rightarrow\{m,M,A,\tan{\beta},M_z,m_t\}$. The last
change of variables introduces a jacobian factor,
\begin{eqnarray}
\label{pdf_new}
p(g_i,m_t, m, M, A, \tan\beta| {\rm data} ) 
= {\cal L}_{\rm rest}\ J|_{\mu=\mu_0}\ p(g_i, y_t, m, M, A, B, \mu=\mu_0)\ \nonumber
\end{eqnarray}
where,
\begin{eqnarray}
\label{jacobiano}
J = \frac{\partial \mu}{\partial M_Z}\;\frac{\partial y_t}{\partial m_t}\;\frac{\partial B}{\partial t\beta}
\simeq \frac{1}{4}(g^2+g^{\prime 2})^{1/2} \left[\frac{E}{R_\mu^2}\right]
\frac{B_{low}}{\mu} \frac{t^2-1}{t(1+t^2)}
\left(\frac{y_t}{y_t^{low}}\right)^2 s_\beta^{-1}.
\end{eqnarray}
In addition, we have developed sensible priors which assume that
soft-breaking term share a common origin. This analysis have been
implemented using \texttt{MultiNest} \cite{Feroz:2007kg} algorithm as
implemented in the \textit{SuperBayeS} \cite{superbayes} code which
incorporate \texttt{SoftSusy} \cite{Allanach:2001kg}, \texttt{SusyBSG}
\cite{Degrassi:2007kj}, \texttt{SuperIso} \cite{Mahmoudi:2008tp},
\texttt{MicrOMEGAs} \cite{micromegas} codes.

As can be seen in Fig
1, besides making the high-energy parameter space quite irrelevant,
the EW breaking has a another remarkable effect, the probability
distribution ($\mathrm{pdfs}$) based on logarithmic or flat prior are
quite similar after the incorporation of the EW scale.
\begin{figure}
\centering
\psfig{figure=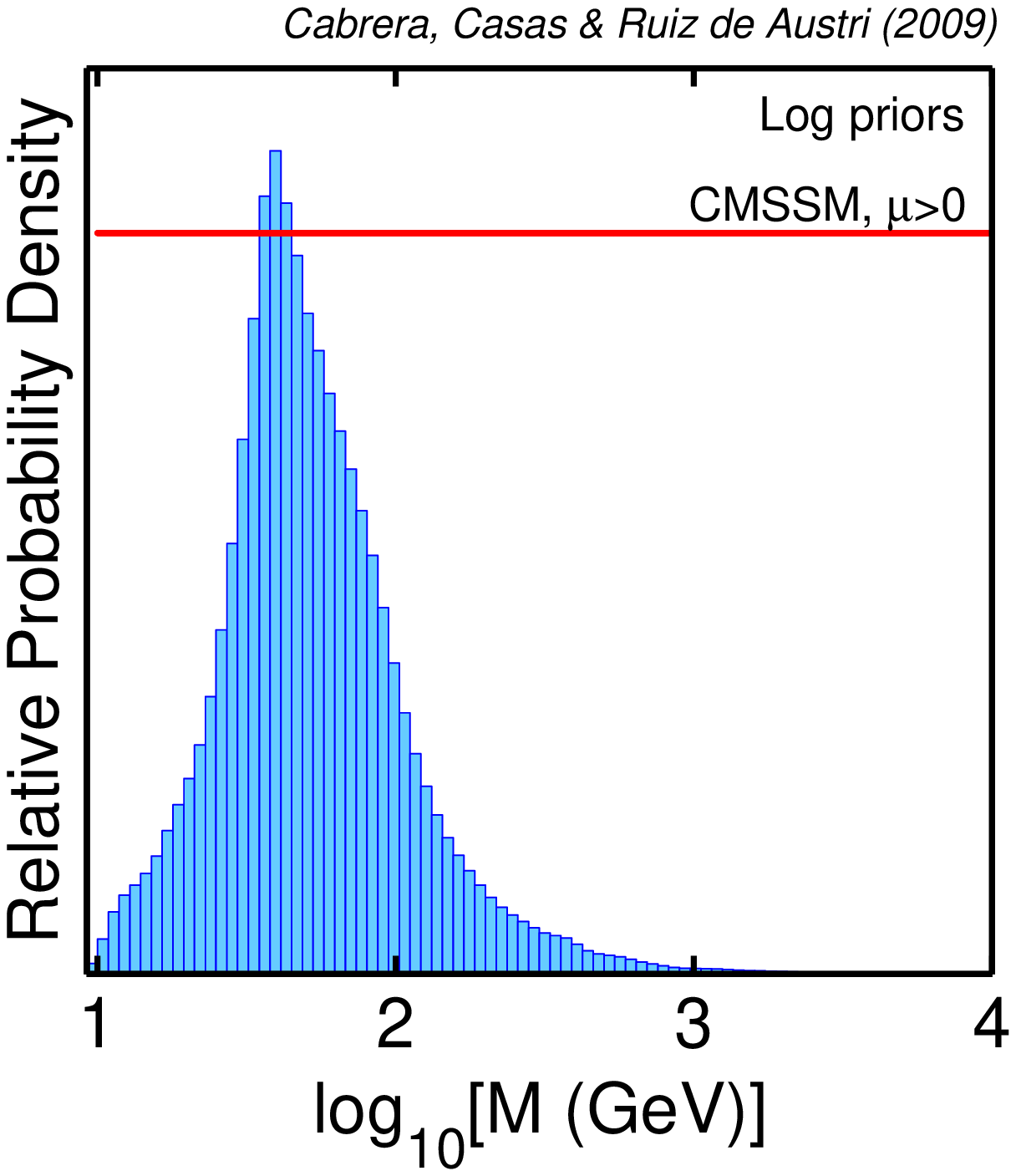,width=1.7in}\hspace{1.0in}
\psfig{figure=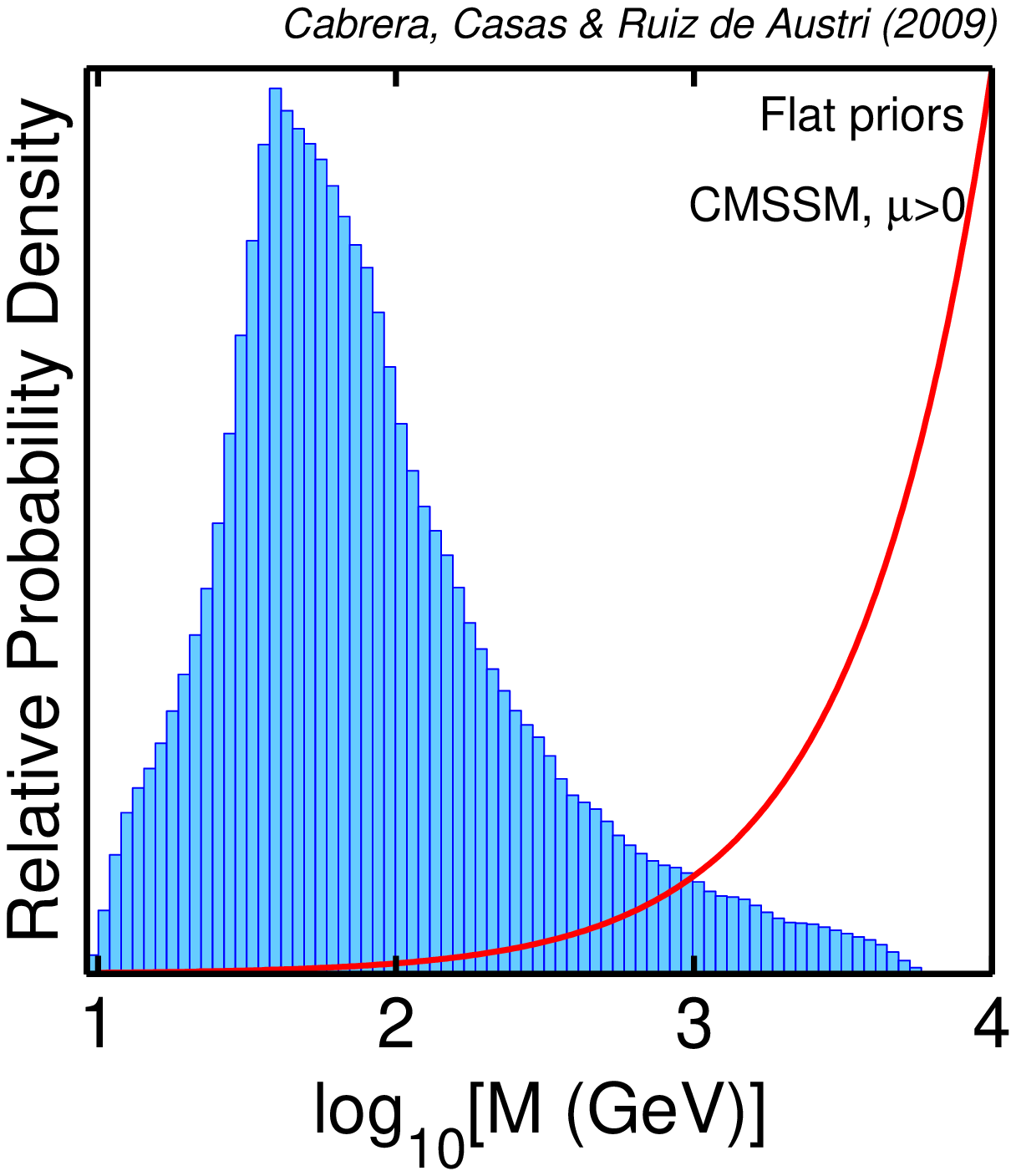,width=1.7in}
\caption{1D marginalized posterior probability distribution of the M
  for logarithmic (left panels) and flat (right panel) priors in the
  $\mu>0$ case, for a scan including the EW breaking information
  ($M_z^{\mathrm{exp}}$). The red lines represents the marginalized
  prior.}
\label{peff_1D}
\end{figure}

\section{Experimental Constraints}

In this section we incorporate all the relevant experimental
information to the likelihood piece of the probability
distribution. We start by considering the most reliable and robust
pieces of experimental information: EW and B(D)-physics observables
and lower bounds on the masses of supersymmetric particles and the
Higgs mass. The left panel of Fig. 2 shows the pdf of $\{M-m\}$ plane
once this experimental information is incorporated. Clearly the bulk
of the probability is now pushed into higher energy. This effect is
basically due to the Higgs mass bound. It is well known that the
tree-level Higgs mass is bounded from above by $M_Z$, so radiative
corrections are needed. Concerning the other observables, everything
works fine as long as SUSY is not at too low scale. We also show the
discovery reach of LHC \cite{Baer:2009dn} for 1 $fb^{-1}$ and 100
$fb^{-1}$. These lines correspond to $A = 0$ and $\tan{\beta}=45$, but
they provide a good indication for the LHC discovery potential.

Next, we have added information about anomalous magnetic momentum of
the muon, $a_\mu$. Taking $e^+ e^- \rightarrow \rm{hadrons}$ data, there
is a $3.3\sigma$ discrepancy between the experimental and SM theoretical
prediction, which has been often claimed as a signal of new physics.
If one accept this, the supersymmetric masses should be brought to
quite small values in order to produce a large enough contribution
$\delta^{\rm{MSSM}}a_{\mu}$ to reconcile theory and experiment, as we
can see in the center panel of Fig. 2. If we instead use $\tau -$decay
data there is no big discrepancy, SUSY contribution does not need to
be large and the pdfs are essentially unchanged by its inclusion.

Supersymmetry offers good candidates for Cold Dark Matter (CDM), the
most popular and natural one is the lightest neutralino.
The right panel of Fig. 2 shows the pdfs after the assumption of
CDM is made of neutralinos and without the information from $a_\mu$. \\
\begin{figure}
\centering
\psfig{figure=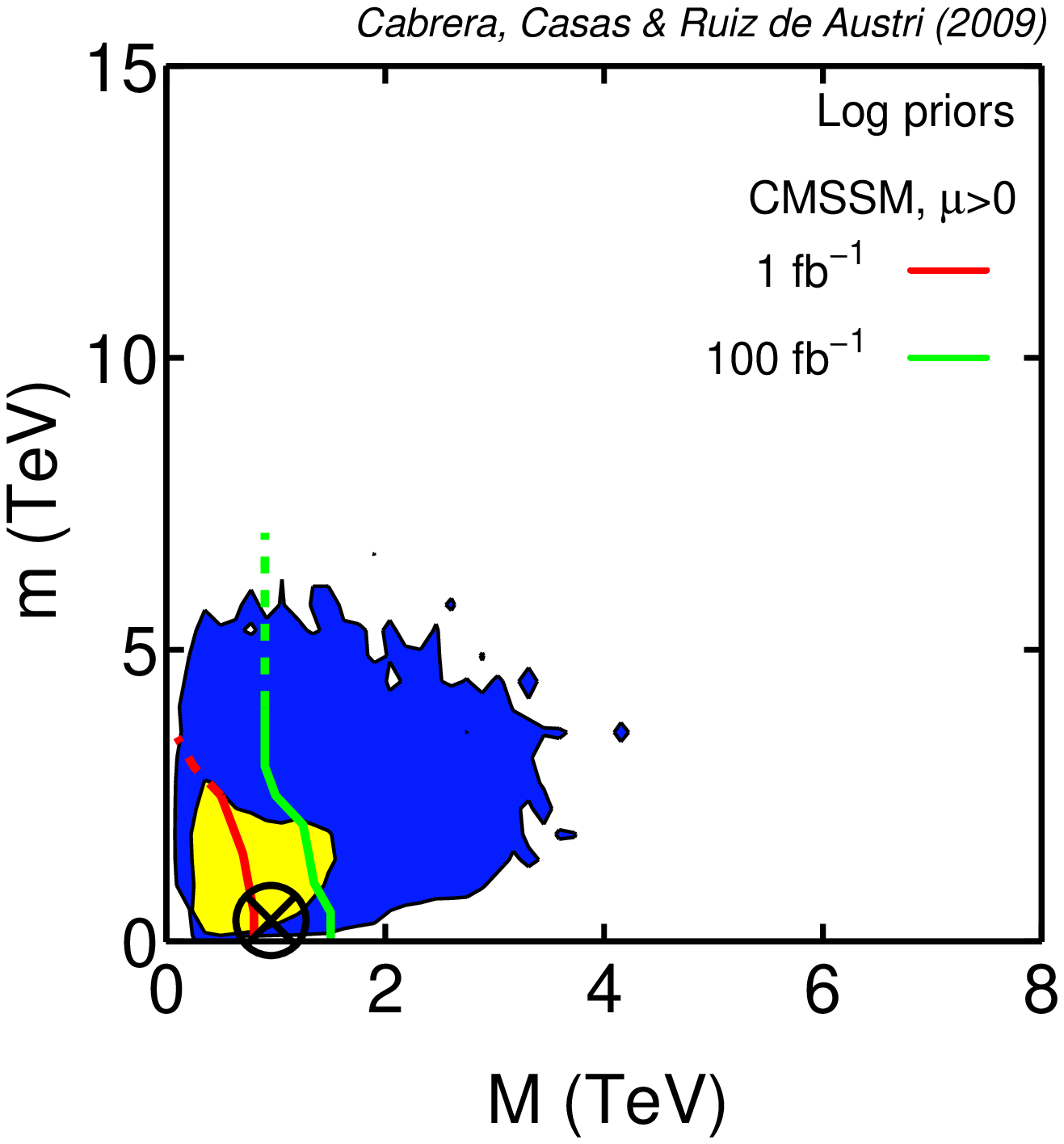,width=1.7in}\hspace{0.4in}
\psfig{figure=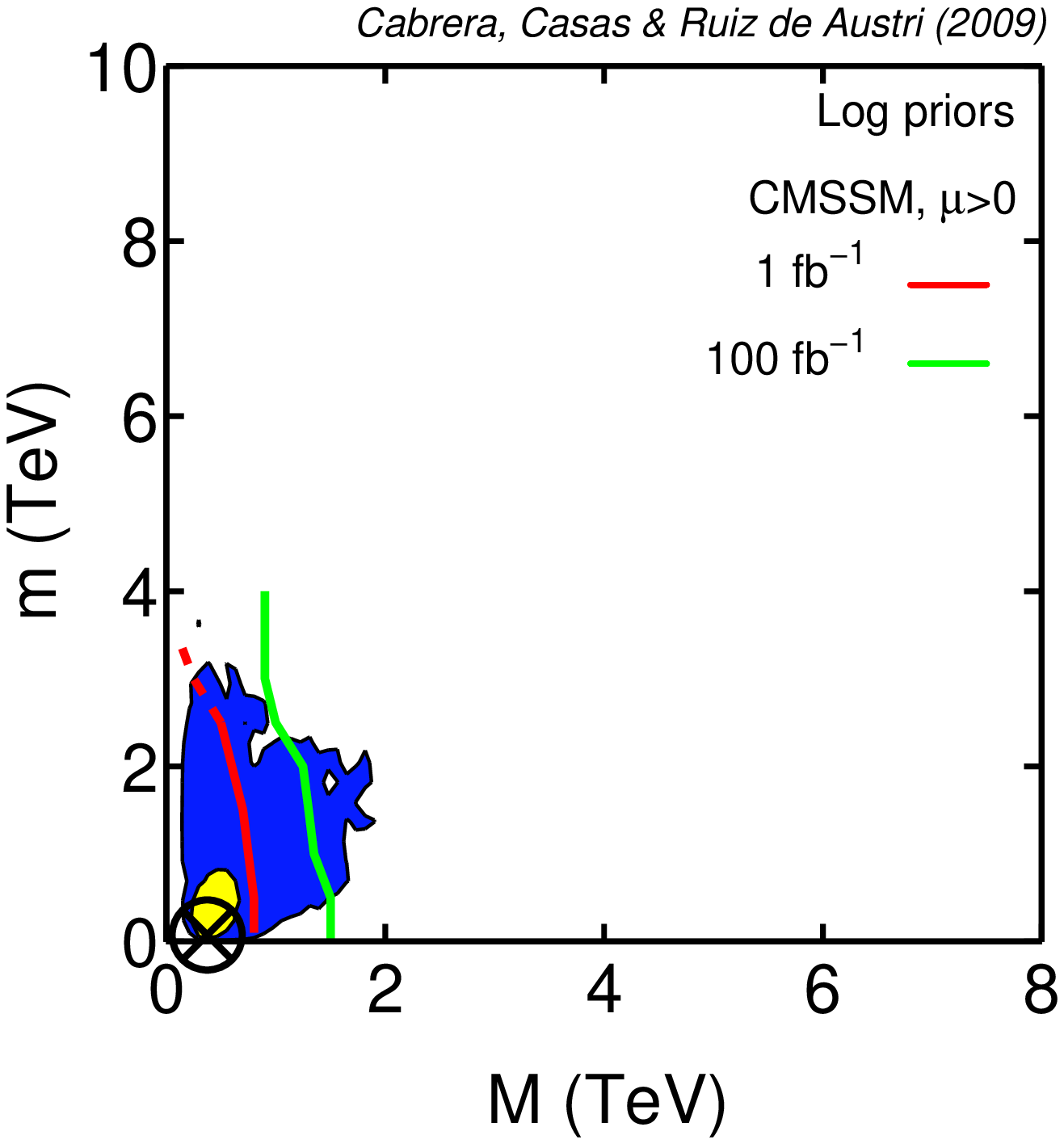,width=1.7in}\hspace{0.4in}
\psfig{figure=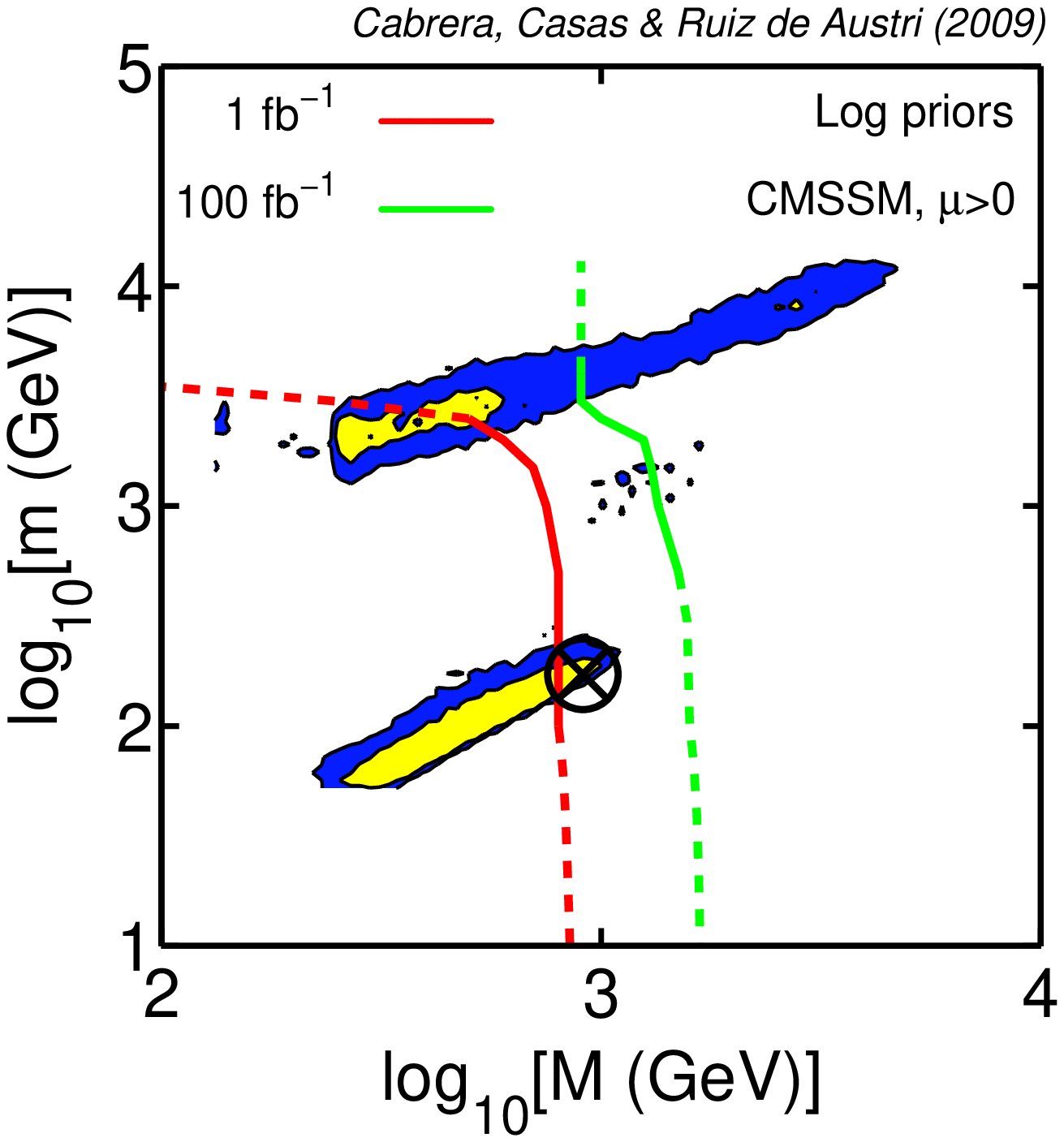,width=1.7in}\\
\psfig{figure=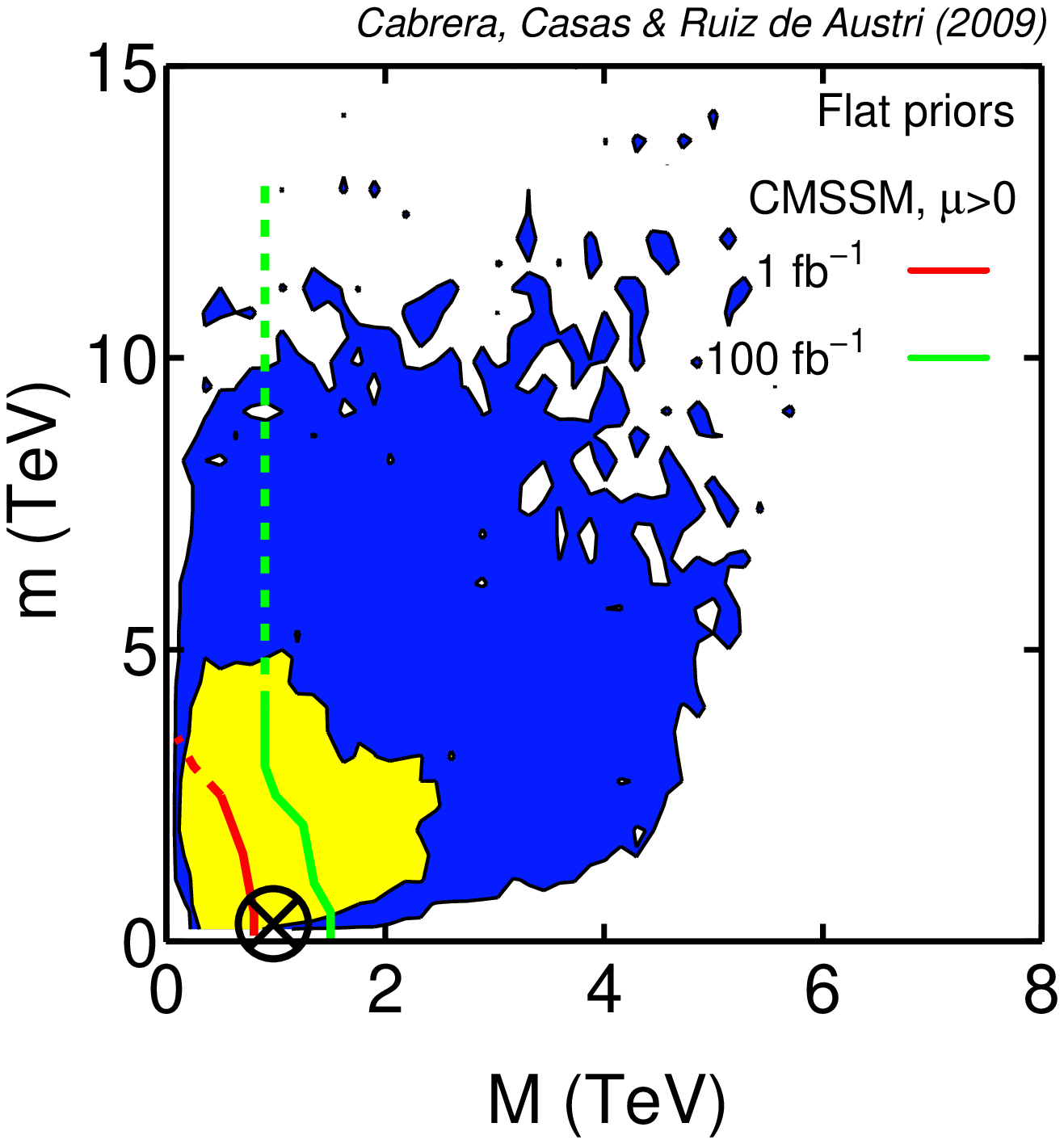,width=1.7in}\hspace{0.4in}
\psfig{figure=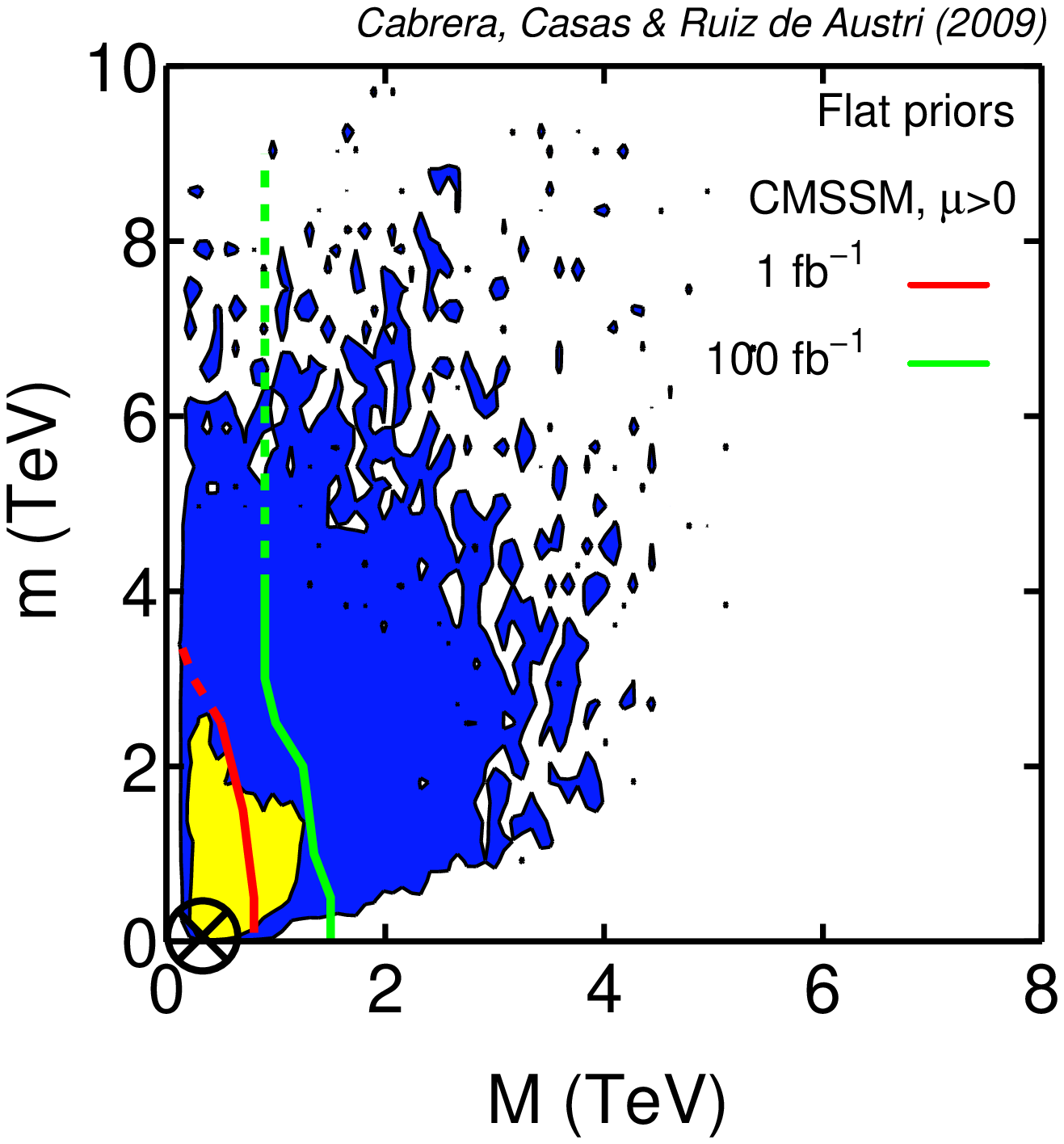,width=1.7in}\hspace{0.4in}
\psfig{figure=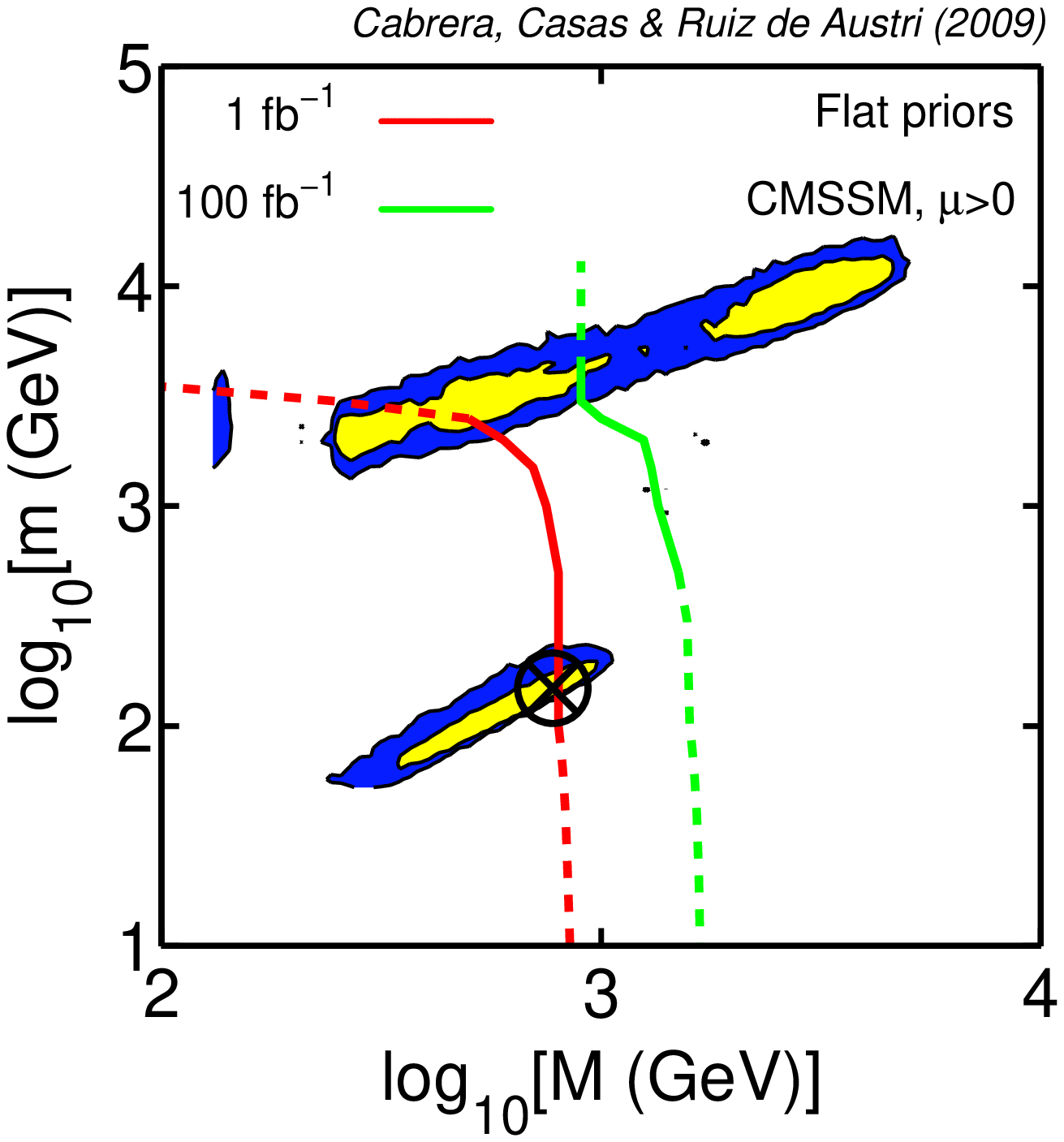,width=1.7in}
\caption{2D marginalized posterior probability distribution for
  logarithmic (upper panel) and flat (lower panel) priors in the
  $\mu>0$ case including: EW + B(D)-physics observables (left panel);
  + $a_{\mu}$ (center panel); + CDM (right panel). The inner and outer
  contours enclose respective $68\%$ and $95\%$ joint regions. }
\label{pdf_2D}
\end{figure}
We have repeated the same analysis for $\mu<0$. We have evaluated the
ratio of the evidence of both cases, the Bayes factor, in order to
compare the relative probability of the $\mu>0$ and $\mu<0$
branches. The $\mu < 0$ is slightly favoured, due to its capability to
reproduce the central value of $b\rightarrow s\,\gamma$, but the
effect is not really significant. 
On the other hand $a_\mu$ favours $\mu>0$ branch, this effect is
stronger when $\Omega_{DM}$ is included. 
This is because $\Omega_{DM}$ constraints favours the low-energy
region of the parameter space, and this is strongly preferred by
$a_{\mu}$.\\
%
%

In conclusion, LHC offers an exciting horizon for SUSY discovery, but
there is still a possibility that this escapes detection, especially
if the Higgs mass is not close to its present experimental bound.

\section*{Acknowledgement}

I thank Alberto Casas and Roberto Ruiz de Austri for the enjoyable
collaboration leading to the work reported here. This work was
supported by the MICINN, Spain, under contract FPA 2007--60252 and by
the Comunidad de Madrid project (HEPHACOS; S2009/ESP-1473). I also
thank the financial support of the CSIC through a predoctoral research
grant (JAEPre 07 00020).


\section*{References}

\begin{thebibliography}{99}
%
\bibitem{Cabrera:2009dm}
  M.~E.~Cabrera, A.~Casas and R.~R.~de Austri,
  arXiv:0911.4686 [hep-ph]. To appear in JHEP
%
\bibitem{Ellis:1986yg}
  J.~R.~Ellis, K.~Enqvist, D.~V.~Nanopoulos and F.~Zwirner,
  Mod.\ Phys.\ Lett.\  A {\bf 1} (1986) 57.
%
\bibitem{Barbieri:1987fn}
  R.~Barbieri and G.~F.~Giudice,
  Nucl.\ Phys.\  B {\bf 306} (1988) 63.
%
\bibitem{Cabrera:2008tj}
  M.~E.~Cabrera, J.~A.~Casas and R.~Ruiz de Austri,
  JHEP {\bf 0903} (2009) 075
  [arXiv:0812.0536 [hep-ph]].
%
\bibitem{Feroz:2007kg}
F. Feroz and  M.~P. Hobson  
Mon. Not. Roy. Astron. Soc. \textbf{384} 449 (2008).
%

\bibitem{superbayes} Available from: \texttt{http://superbayes.org}

\bibitem{Allanach:2001kg}
  B.~C.~Allanach,
  Comput. \ Phys. \ Commun. {\bf 143} (2002) 305  
  [arXiv:hep-ph/0104145].
%
\bibitem{Degrassi:2007kj}
G.~Degrassi, P.~Gambino and P.~Slavich,
Comput.\ Phys.\ Commun.\  {\bf 179} (2008) 759
[arXiv:0712.3265 [hep-ph]].

\bibitem{Mahmoudi:2008tp}
F.~Mahmoudi,
Comput.\ Phys.\ Commun.\  {\bf 180}, 1579 (2009)
[arXiv:0808.3144 [hep-ph]].
%
%
\bibitem{micromegas}
G.~Belanger, F.~Boudjema, A.~Pukhov and A.~Semenov,
Comput.\ Phys.\ Commun.\  {\bf 149} (2002) 103 [hep-ph/0112278]; 
Comput.\ Phys.\ Commun.\  {\bf 174}, 577 (2006) [hep-ph/0405253].
%
\bibitem{Baer:2009dn}
H.~Baer, V.~Barger, A.~Lessa and X.~Tata,
JHEP {\bf 0909} (2009) 063
[arXiv:0907.1922 [hep-ph]].
%
\end{thebibliography}
\end{document}